\documentclass{article}
\usepackage[margin=1in]{geometry}
\usepackage{slashed}
\usepackage{feynmp}
\usepackage[numbers,sort&compress]{natbib}
\usepackage{authblk} 
\usepackage{fancyhdr} 

\def\nn{\nonumber}

\title{Basis invariant description of chemical equilibrium with implications for a recent axionic leptogenesis model }
\author{Bowen Shi}
\author{Stuart Raby}
\affil{\emph{Department of Physics}\\\emph{The Ohio State University}\\\emph{191 W.~Woodruff Ave, Columbus, OH 43210, USA}}

\begin{document}

\rhead{OHSTPY-HEP-T-15-004}
\renewcommand\headrule{} 

\maketitle
\abstract{We provide a systematic  treatment of  chemical equilibrium in the presence of a specific type of time dependent background. The type of time dependent background we consider appears, for example, in recently proposed axion/Majoron leptogenesis models \cite{Kusenko:2014uta,Ibe:2015nfa}.  In describing the chemical equilibrium we use quantities which are invariant under redefinition of fermion phases (we refer to this redefinition as a change of basis for short\footnote{In this paper, change of basis does not mean change of Lorentz frame. All calculations in this paper are performed in the center-of-momentum frame of the thermal plasma, i.e. the Lorentz frame in which the average momentum of particles is zero.}), and therefore it is a basis invariant treatment. The change of the anomaly terms due to the change of the path integral measure \cite{Fujikawa:1979ay,Fujikawa:1983bg} under a basis change is taken into account. We find it is useful to go back and forth between different bases, and there are insights which can be more easily obtained in one basis rather than another. A toy model is provided to illustrate the ideas.

For the axion leptogenesis model \cite{Kusenko:2014uta}, our result suggests that at $T > 10^{13}$ GeV, when sphaleron processes decouple and $\Gamma_{B+L} \ll H < \Gamma_L$ (where $H$ is the Hubble parameter at temperature $T$ and $\Gamma_L$ is the $\Delta L = 2$ lepton number violating interaction rate) , the amount of $B-L$ created is controlled by the smallness of the sphaleron interaction rate, $\Gamma_{B+L}$. Therefore it is not as efficient as described. In addition, we notice an interesting modification of gauge boson dispersion relations at subleading order.}


\section{Introduction}
Recently, novel models  of leptogenesis were proposed  \cite{Kusenko:2014uta,Ibe:2015nfa} which employ the idea of spontaneous baryogenesis pioneered by Cohen and Kaplan \cite{Cohen:1987vi}. The key idea is the existence of a specific type of time dependent classical background field in the early universe. In \cite{Kusenko:2014uta} the background  field comes from an axion which couples to the electroweak gauge fields $W^a$  and $B^a$.\footnote{In this paper we use $a,b,c,d$ as space-time indices.}  The axion is assumed to get nonzero mass from coupling to hidden sectors.\footnote{Unlike the QCD axion, electroweak axion could not generate a mass by anomaly. Also, for the purpose of leptogenesis, the mass needs to be large.} In \cite{Ibe:2015nfa} the time dependent background comes from a Majoron which is assumed to get mass from new physics at the Planck scale. In the early universe, after inflation a homogeneous background is produced which, in general, does not lie at its minimum, assuming the corresponding symmetry is broken before the end of inflation. When either the axion or Majoron run down their respective potentials at a temperature scale $T$ satisfying $H(T)\sim  m(T)$ (where $m$ is the mass of the background axion or Majoron field), a homogeneous, time dependent background field is produced. These models illustrate the interesting possibility of explaining the observed baryon asymmetry $\eta_B^0\simeq 6\times 10^{-10}$ \cite{Ade:2013zuv} in a $CPT$ violating background field configuration without using the $CP$ violation in the fundamental theory ($CPT$ is assumed to be a good symmetry of the fundamental theory).

While the time dependent background field (which may be considered as a coherent state with zero momentum) is not in thermal equilibrium, nonzero lepton number or baryon number can be generated when the lepton/baryon number violating interactions are in equilibrium, i.e. the interaction rates are large compared to the Hubble parameter, $H$. It is not always necessary for the system to reach the equilibrium value, and when the system evolves  towards the equilibrium value with nonzero baryon/lepton number, a nonzero baryon/lepton number asymmetry is generated. In the case the equilibrium value is not reached, the amount of asymmetry produced is determined by the relevant interaction rates which enter the Boltzmann equations.\\

In this paper we discuss the change of basis invariance of physics, which is relevant for the axion/Majoron leptogenesis models.  In particular, we work out the equilibrium values of B and L. The change of anomaly terms, due to the change of the path integral measure \cite{Fujikawa:1979ay,Fujikawa:1983bg} under basis changes, is taken into account, and therefore, our discussion should be distinguished from the Appendix of \cite{Ibe:2015nfa} where the basis changes are discussed in the context of a classical Lagrangian.

To the best of our knowledge, we are the first to  provide a systematic  basis invariant treatment of chemical equilibrium in such a time dependent background.\footnote{For a general time dependent background, both kinetic and chemical equilibrium do not exist, but the specific type of time dependent background we consider allows both kinetic and chemical equilibrium.} For the purpose of obtaining a basis invariant treatment, we use quantities which are invariant under the basis changes we consider, like the fermion number density $n$, the fermion occupancy $f(\vec{p})$, and the fermion {\em effective} chemical potential $\bar{\mu}$. On the other hand, the fermion energy $E$ and fermion chemical potential $\mu$ are not invariant under basis changes, and they do not enter our final results. We find insights which are better seen in one basis rather than another, and it is useful to go back and forth between different bases.

A toy model is provided to illustrate most of the ideas. In the toy model we illustrate a simple example of the time dependent background we consider, the type of basis changes we consider and the change of path integral measure under basis changes. By choosing a suitable basis, the Lagrangian becomes time independent and this explains why thermal equilibrium and chemical equilibrium could exist in the type of time dependent background being considered. Quantities which are invariant under the basis changes are discussed, and the chemical equilibrium is described using the invariant quantities (especially the {\em effective} chemical potential $\bar\mu$). The description in different bases are explained at the level of the Boltzmann equation and insight from different bases are discussed.

When applied to the axion leptogenesis model \cite{Kusenko:2014uta}, our result suggests a different equilibrium point than that shown in \cite{Kusenko:2014uta}. Our result shows that $B$ must be generated at the same time $B-L$ is generated, otherwise $B=L=0$. At $T > 10^{13}$ GeV, the $\Delta L=2$ interaction rate per particle satisfies $\Gamma_L > H$.  However the sphaleron interaction rate per particle, $\Gamma_{B+L} \approx 250 \ \alpha_W^5 T \ll H$ \cite{Davidson:2008bu} and it is thus not as effective.  We show that in this limit the amount of $B-L$ created is controlled by the smallness of the sphaleron interaction rate per particle, $\Gamma_{B+L}$, rather than, $\Gamma_L$, and the creation of $B-L$ is not as efficient as described in \cite{Kusenko:2014uta}.  We also show that the end results obtained by the authors in \cite{Ibe:2015nfa} are unchanged;  however the derivation of the effective action was incomplete.  As an aside, we notice a modification of the gauge boson dispersion relation at subleading order which exists in the axion leptogenesis model \cite{Kusenko:2014uta}, but not in the Majoron leptogenesis model \cite{Ibe:2015nfa}.

\section{Change of  basis and the invariance of physics -- a toy model }
Invariance of physics under frame or basis changes plays a key role in modern theoretical physics; such as Lorentz invariance in special relativity, general coordinate invariance in general relativity and gauge invariance in gauge theories.

Changing of fermion phases is central in Fujikawa's way \cite{Fujikawa:1979ay} of understanding quantum anomalies. Here we continue the story of changing fermion phases (we call it change of basis for short) and investigate its implication in thermal dynamics, especially in chemical equilibrium. We find for the type of basis changes we are interested in, the energy or chemical potential of fermions are not invariant quantities. Nevertheless, particle number density $n$, occupancy $f(\vec{p})$, 3-momentum of fermions, {\em effective} chemical potential of fermions, 4-momentum of bosons, the chemical potential of bosons and the dispersion relation of bosons are invariant quantities. By changing basis, a good amount of information can be obtained.\\
\subsection{The toy model in basis (A)}\label{11}
We illustrate the idea using a toy model. Consider four left-handed Weyl fermions  $q_1$,$q_2$, $q_3$ and $l$ in the fundamental representation of an $SU(2)$ gauge group and $W_{ab}$ is the field strength of the $SU(2)$ gauge field with the Lagrangian given by
\begin{equation}
(A)\qquad\mathcal{L}=l^\dagger i\bar{\sigma}^a D_a l+ \sum_{i=1}^3 q_i^\dagger i\bar{\sigma}^aD_a q_i-\frac{1}{2g^2}tr(W_{ab}W^{ab})-\frac{\theta(x)}{16\pi^2}tr(W_{ab}\tilde{W}^{ab})\label{Action}
\end{equation}
where
\begin{equation}
D_a=\partial_a+iW_a .
\end{equation}
We are interested in a homogeneous and time dependent background, so consider
\begin{equation}
\partial_a\theta=(\dot{\theta},0,0,0)=(\delta,0,0,0).
\end{equation}
Here, we consider $\delta=\textrm{const}$ since in the realistic models we will be interested in, $\dot{\theta}$ is slowly changing, and it could be treated as a constant during a period of time when some relevant interactions happen. We have chosen our notation to indicate the similarity between this toy model and the $SU(2)_L$ weak interaction in the standard model (SM).

According to the theorem of global anomaly by Witten \cite{Witten:1982fp}, for an $SU(2)$ gauge theory to be consistent, there must be an even number of $SU(2)$ Weyl fermion doublets (in the fundamental representation) assuming no fermions in other representations of $SU(2)$. There are alternative proofs of Witten's theorem  using Abelian anomaly \cite{deAlwis:1985uj} or non-Abelian anomaly \cite{Elitzur:1984kr,Klinkhamer:1990eb}. Therefore, it is possible to choose a slightly simpler toy model with just two Weyl fermion doublets. However, in the case of two Weyl fermion doublets, the anomalous one-instanton effect \cite{'tHooft:1976up,'tHooft:1976fv} or sphaleron effect \cite{Klinkhamer:1984di,Kuzmin:1985mm,Arnold:1987mh,Arnold:1987zg} involves only two fermions and induces a correction to the fermion propagators which is equivalent to a mass term rather than an interaction. If we want the anomaly to induce an interaction, rather than a mass term at the lowest order, our toy model is the minimal set up.
\subsection{Change of basis from (A) to (B)}
We consider the following change of basis (although change of basis could be more general)
\begin{eqnarray}
&&l\to e^{ic_1\theta(x)}l   \nn\\
&&q_i\to e^{ic_2\theta(x)}q_i\qquad\qquad i=1,2,3   \label{c2}\\
&&c_1+3c_2=1    \nn.
\end{eqnarray}
Fujikawa's method \cite{Fujikawa:1979ay,Fujikawa:1983bg} (especially Ref. \cite{Fujikawa:1983bg})   is very helpful in understanding how an anomaly term changes under fermion phase rotations in chiral gauge theories. Fujikawa's idea is to consider the path integral of the theory; when the phases of the fermions are rotated, the path integral measure of fermions may not be invariant (depending on what rotation is performed and how fermions couple to the gauge fields). This effect is equivalent to adding a term into the classical Lagrangian after the fermion phase rotation. Some useful results of Fujikawa's method are summarized in Appendix \ref{Fujikawa} using our notation. \\
The effect of the rotation Eq. (\ref{c2}) is the following:\\
(1) The change of basis induces a change of the anomaly term (due to the change of path integral measure)
\begin{equation}
\delta\mathcal{L}_{anomaly}=\frac{(c_1+3c_2)\theta(x)}{16\pi^2}\,tr(W_{ab}\tilde{W}^{ab})=\frac{\theta(x)}{16\pi^2}\,tr(W_{ab}\tilde{W}^{ab}).
\end{equation}
Therefore, the original anomaly term is canceled in this new basis.\\
(2) Since $\partial_a\theta=(\delta,0,0,0)$,
the change of basis will also introduce the following terms into the Lagrangian
\begin{equation}
\mathcal{L}_\delta=-c_1\delta l^\dagger l-\sum_{i=1}^3c_2\delta q^\dagger_i q_i .
\end{equation}

Therefore, with a change from basis (A) to  basis (B), the Lagrangian becomes
\begin{equation}
(B)\qquad\mathcal{L}'=l^\dagger i\bar{\sigma}^a D_a l-\delta_l l^\dagger l+ \sum_{i=1}^3 (q_i^\dagger i\bar{\sigma}^aD_a q_i -\delta_q q_i^\dagger q_i) -\frac{1}{2g^2}tr(W_{ab}W^{ab}),
\end{equation}
where
\begin{equation}
\delta_l+3\delta_q=\delta\qquad\qquad \delta_l=c_1\delta\qquad\qquad \delta_q=c_2\delta .
\end{equation}
\subsection{Energy shift and effective chemical potential}
\subsubsection{Energy shift}
The terms $\delta_l l^\dagger l$ and $\delta_q q_i^\dagger q_i$ give energy shifts to $l$ and $q_i$ particles.\footnote{Since we have freedom to choose $c_1$ and $c_2$ keeping $c_1+3c_2=1$, we are actually considering a lot of possible basis changes parametrized by a real number. Each of them gives you  different energy shifts. } At the classical level, this may be seen by considering classical solutions of the free part of the Lagrangian (here, free means getting rid of  interactions). The reason we only consider the free part of the Lagrangian is the assumption  that for a weakly interacting plasma, particle can be defined using the free part of the Lagrangian. For example, consider the $l(x)$ field
\begin{equation}
\mathcal{L}_0=l^\dagger i\bar{\sigma}^a\partial_al\qquad\Longrightarrow\qquad l=u(p)e^{-ipx}\qquad \textrm{or}\qquad l=v(p)e^{ipx}\label{L1}
\end{equation}
where $p^a=(\vert\vec{p}\vert, \vec{p})$.
After basis changes $l\to e^{i\delta_l t}l$
\begin{equation}
\mathcal{L}'_0=l^\dagger i\bar{\sigma}^a\partial_al-\delta_l l^\dagger l\qquad\Longrightarrow\qquad l=u(p)e^{-ipx}e^{-i\delta_l t}\qquad \textrm{or}\qquad l=v(p)e^{ipx}e^{-i\delta_l t}\label{L2}.
\end{equation}
The solutions with factor $e^{-ipx}$ are called particle solutions, and the solutions with factor $e^{ipx}$ are called antiparticle solutions.
The energy of particle and antiparticle is shifted in opposite directions.
\begin{eqnarray}
E_l(\vec{p})&=&\vert\vec{p}\vert+\delta_l \\
E_{\bar{l}}(\vec{p})&=&\vert\vec{p}\vert-\delta_l \nn\\
E_{q_i}(\vec{p})&=&\vert\vec{p}\vert+\delta_q \nn\\
E_{\bar{q}_i}(\vec{p})&=&\vert\vec{p}\vert-\delta_q \nn
\end{eqnarray}
Energy shifts are not only defined in basis (B) but also defined in basis (A). The amount of energy shift in basis (A) is zero. More details about energy shifts may be found in Appendix \ref{8}.
\subsubsection{Effective chemical potential}
We have seen that energy shifts may come about in our toy model by basis changes.
In systems with energy shifts, it is convenient to define an effective chemical potential $\bar{\mu}$.

Let us consider the following shifts of particle energy, and mass $m$ is added for a general definition. In the toy model and the relevant temperature scales of the realistic models in Sec. \ref{realistic}, the particles are massless.

\begin{equation}
E(\vec{p})= E_0(\vec{p})+\delta\qquad\qquad\textrm{with}\qquad\qquad E_0(\vec{p})=\sqrt{\vec{p}^2+m^2}
\end{equation}
In kinetic equilibrium the occupancy is
\begin{equation}
f(\vec{p})=\frac{1}{e^{\frac{E(\vec{p})-\mu}{T}}\pm1}=\frac{1}{e^{\frac{E_0(\vec{p})+\delta-\mu}{T}}\pm1}.
\end{equation}
For the purpose of calculating the occupancy it is convenient to define the effective chemical potential
\begin{equation}
\bar{\mu}\equiv \mu-\delta,
\end{equation}
then
\begin{equation}
f(\vec{p})=\frac{1}{e^{\frac{E_0(\vec{p})-\bar{\mu}}{T}}\pm1}.
\end{equation}
For each internal degree of freedom, the number density is
\begin{equation}
n=\int \frac{d^3p}{(2\pi)^3}f(\vec{p}).
\end{equation}
In a system with chemical potential $\mu$ and energy shift $\delta$, the occupancy and number density can be calculated as if there is no energy shift and with the {\em effective} chemical potential $\bar{\mu}$.
\subsection{Invariant quantities of the toy model}\label{invariant}
For a basis invariant description of physics, it is very important to find  invariant quantities\footnote{Quantities which change according to some simple rules (under basis changes) are called covariant quantities. Covariant quantities can be important also. For example, in general relativity,  vector and tensor are important covariant quantities.} under basis changes. For the toy model, there are fermions and gauge bosons. In this section, we will discuss the invariant quantities of fermions and gauge bosons which provide an invariant description of the system.
\subsubsection{Invariant quantities for fermions}
We have seen that the fermion energy gets shifted and the shift does not take the same value in basis (A) and basis (B). On the other hand, from the requirement that physics is independent of basis, the number of fermions with some specific 3-momentum $\vec{p}$ are the same from the viewpoint of both bases. At the classical level, it may be seen by considering a classical solution of the free part of the Lagrangian (\ref{L1}), (\ref{L2}).
Under basis change $l\to e^{i\delta_l t}l$, the  particle solution
\begin{equation}
l=u(p)e^{-ipx}\quad\longrightarrow\quad l=u(p)e^{-ipx}e^{-i\delta_l t}
\end{equation}
The physics requirement is that, a particle described by the solution $u(p)e^{-ipx}$ will become a particle described by the solution $u(p)e^{-ipx}e^{-i\delta_l t}$ (with the same 3-momentum $\vec{p}$) after the basis change $l\to e^{i\delta_l t}l$. Similarly for antiparticles. Therefore, the number of fermions with some specific 3-momentum $\vec{p}$ should be the same from the viewpoint of both bases (for a further discussion of this point see Sec. \ref{sec:equilibrium}). This means that the occupancy $f(\vec{p})$ is an invariant quantity.\\
For each internal degree of freedom
\begin{equation}
n=\int \frac{d^3p}{(2\pi)^3}f(\vec{p}).
\end{equation}
Therefore, $f(\vec{p})$ invariant implies the number density $n$ is an invariant quantity.\\
For fermions, in kinetic equilibrium we have
\begin{equation}
f(\vec{p})=\frac{1}{e^{\frac{E_0(\vec{p})-\bar{\mu}}{T}}+1}.
\end{equation}
Here, $E_0(\vec{p})$ is an invariant, therefore, the effective chemical potential $\bar{\mu}$ is invariant. On the other hand, the chemical potential $\mu=\bar{\mu}+\delta$ is not invariant since $\delta$ is not invariant.

 To summarize, for fermions, the 3-momentum,  the occupancy $f(\vec{p})$, the number density $n$, and (when the system is in kinetic equilibrium) the effective chemical potential $\bar{\mu}$ are invariant quantities.
\subsubsection{Invariant quantities of gauge bosons}
In the change of basis  considered in the toy model, we did not transform the gauge field. Therefore, we expect that everything about the gauge boson in basis (A) and (B) is the same. To be specific, we expect the 4-momentum, the occupancy $f(\vec{p})$,  the number density $n$, the chemical potential and the dispersion relation of the gauge boson to be invariant. We find there is an interesting subtlety concerning the dispersion relation of the gauge boson, and it conforms to our expectation.\\

In the context of an Abelian gauge theory, consider
\begin{equation}
S=\int d^4x-[\frac{1}{g'^2}B_{ab}B^{ab}+\frac{\theta(x)Y^2}{16\pi^2}B_{ab}\tilde{B}^{ab}]\qquad\qquad \partial_a\theta=(\delta,0,0,0).
\end{equation}
A similar theory has been considered by Carroll, Field and Jackiw \cite{Carroll:1989vb} in the context of electrodynamics modified by a Lorentz-violating Chern-Simons term $\mathcal{L}_{CS}=-p_aA_b\tilde{F}^{ab}$. [The term $\theta F\tilde{F}$ is equivalent to $-2(\partial_a\theta) A_b\tilde{F}^{ab}$ up to a total derivative.] By solving the classical equations of motion it can be shown that the dispersion relation of the gauge boson is modified:
\begin{equation}
\partial_a B^{ab}=b_a\tilde{B}^{ab}\qquad b_a=(-\frac{g'^2Y^2}{16\pi^2}\delta,0,0,0) \qquad\Longrightarrow\qquad \omega^2=k^2\pm \frac{g'^2 Y^2}{16\pi^2}k\delta
\end{equation}
with $\pm$ for the two possible circularly polarized modes, and $k\equiv \vert \vec{k}\vert$.\\
For our non-Abelian gauge field in the toy model, from the view point of basis (A), after neglecting the nonlinear terms, we expect a similar modification of the dispersion relation. For the action
\begin{equation}
S=\int d^4x-[\frac{1}{2g^2}tr(W_{ab}W^{ab})+\frac{\theta(x)}{16\pi^2}tr(W_{ab}\tilde{W}^{ab})]\qquad\qquad \partial_a\theta=(\delta,0,0,0),
\end{equation}
the dispersion relation would be
\begin{equation}
\omega^2=k^2\pm \frac{g^2}{8\pi^2}k\delta \label{dispersion}.
\end{equation}
From the requirement that physics is independent of basis, we expect the same dispersion relation in basis (B). But in basis (B) there is no $\theta(x)W\tilde{W}$ term, and how should the dispersion relation of the gauge boson be modified? It may not be too surprising that from the point of view of basis (B), the same modification of the dispersion relation comes from a fermionic 1-loop correction to the propagator of the gauge boson.\footnote{The terms $\delta_l l^\dagger l$ and $\delta_q q_i^\dagger q_i$ which cause energy shifts of the fermions also contribute to the loop diagram, and this makes it possible to modify the dispersion relation of the boson.} A calculation of relevant 1-loop correction in the context of QED (set fermion mass $m=0$) may be found in \cite{PerezVictoria:1999uh}, and we also notice that in their paper the result was explained using the idea of basis changes  taking into account the change of anomaly term from the path integral measure. From the viewpoint of basis (B), the fermions get energy shifts of order $\delta$, and the gauge coupling $g^2$ enters into the boson dispersion relation  due to the loop. This result justifies our basis independent argument for gauge bosons.\\
Summary and comments:

(1) The basis invariant quantities for gauge bosons are: the 4-momentum, the occupancy $f(\vec{p})$, the number density $n$, the chemical potential and the dispersion relation.

(2) The dispersion relation  of the gauge boson is given by $\omega^2=k^2\pm \frac{g^2}{8\pi^2}k\delta$ [as is shown in Eq. (\ref{dispersion})]. Note that the second term is linear in $k$, and therefore this modification is not a mass term; this effect is special for time dependent axion background and it is a zero temperature effect. When $k\gg\delta$, this is equivalent to an energy shift $\omega\simeq k\pm \frac{g^2}{16\pi^2}\delta$. Since the energy shift of the gauge boson  is suppressed by a factor $\frac{g^2}{16\pi^2}$, compared to the energy shift of the fermions, we will neglect this small energy shift for the gauge boson in considering the equilibrium (assuming $g^2\ll 1$). In chemical equilibrium, we will use $\mu_W=0$, and as we neglect the small energy shift, we will not use an effective chemical potential for gauge bosons since it is equal to the chemical potential in every basis, $\bar{\mu}_W\simeq \mu_W$.

(3) The dispersion relation  of the gauge boson, $\omega^2=k^2\pm \frac{g^2}{8\pi^2}k\delta$ [as is shown in Eq. (\ref{dispersion})] has an instability at small momentum, $k<\frac{g^2}{8\pi^2}\delta$. When the thermal mass of  gauge boson $m\sim gT$ (see for example \cite{Braaten:1990ee}) is taken into account, the instability no longer exists (assuming both $T\gg \delta$ and $g\ll1$). In fact, for a thermal averaged momentum $k\sim T$, the thermal correction to the frequency is of the order $g^2 T$ \footnote{The thermal correction treats the two circularly polarized (transverse) modes in the same way, while the axion background distinguishes the two modes. In thermal plasma there are longitudinal modes also, but only for momentum smaller than $gT$, and therefore not relevant here. }, and the correction due to the axion background to the frequency is of the order $\pm g^2\delta$. Both  are second order in the gauge coupling $g$, and when $T\gg \delta$, the thermal correction is  bigger then the axion correction. Nevertheless, the fact that the axion background correction treats $\pm$ circularly polarized modes (or helicity) differently may have interesting consequences, see Sec. \ref{sub-leading observable} for more detail.

\subsection{Chemical equilibrium \label{sec:equilibrium}}
Generally speaking, chemical equilibrium does not exist in systems with a time dependent Lagrangian. Our toy model in basis (A) is time dependent, and at first sight, it is not clear whether it is possible to have chemical equilibrium. Nevertheless, in basis (B), the Lagrangian is time independent, and it is possible to define chemical equilibrium. We first provide a treatment of chemical equilibrium in basis (B), and then use the invariant quantities to obtain the equilibrium in basis (A). The viewpoint from different bases are discussed at the level of Boltzmann equation.
\subsubsection{A brief review of chemical equilibrium}
As we will use chemical equilibrium in a nontrivial way, it is worthwhile to briefly review it here, together with the derivation from the Boltzmann equation. \\
For a process
\begin{equation}
A+B+\ldots \rightleftharpoons C+D+\ldots
\end{equation}
Start with the Boltzmann equation\footnote{If there are identical particles in the interaction, the phase space needs to be modified.} which may be found in \cite{Kolb:1990vq} (here I set the Hubble parameter  $H=0$)
\begin{eqnarray}
\frac{dn_A}{dt}&\equiv&-\int  d\Pi_Ad\Pi_B\ldots d\Pi_C d\Pi_D\ldots\nonumber\\
&&\times (2\pi)^4\delta^4(p_A+p_B\ldots -p_C-p_D\ldots)\nonumber\\
&&\times \left[\begin{array}{l}\,\,\,\,\vert\mathcal{M}\vert^2_{A+B+\cdots\to C+D+\cdots}f_A f_B\cdots (1\pm f_C)(1\pm f_D)\cdots\\-\vert\mathcal{M}\vert^2_{C+D+\cdots\to A+B+\cdots} f_Cf_D\cdots (1\pm f_A)(1\pm f_B)\cdots\end{array}\right]
\end{eqnarray}
where for each internal degree of freedom $d\Pi=\frac{d^3p}{(2\pi)^3}$ is the phase space factor,  $f(\vec{p})$ is the occupancy and in kinetic equilibrium $f(\vec{p})=\frac{1}{\exp{\frac{E(\vec{p})-\mu}{T}} \pm1}$ .\\
(1) If we have
\begin{equation}
\vert\mathcal{M}\vert^2_{A+B+\ldots\to C+D+\ldots}=\vert\mathcal{M}\vert^2_{C+D+\ldots\to A+B+\ldots}\label{flat1}
\end{equation}
then, in chemical equilibrium we could derive
\begin{equation}
\mu_A+\mu_B+\ldots=\mu_C+\mu_D+\ldots \label{flat2}
\end{equation}
(2) If $T$ is violated, we may have
\begin{equation}
\vert\mathcal{M}\vert^2_{A+B+\ldots\to C+D+\ldots}\ne\vert\mathcal{M}\vert^2_{C+D+\ldots\to A+B+\ldots}
\end{equation}
then, in chemical equilibrium one would derive
\begin{equation}
\mu_A+\mu_B+\ldots=\mu_C+\mu_D+\ldots+T\,\ln\bigg\vert\frac{\mathcal{M}_{C+D+\ldots\to A+B+\ldots}}{\mathcal{M}_{A+B+\ldots\to C+D+\ldots}}\bigg\vert^2 \label{7}
\end{equation}
Throughout this paper, we do not need to worry about the situation in Eq. (\ref{7}) because we neglect the small $CP$ nonconservation in the weak interactions when considering the axionic/Majoron leptogenesis models.\\
(3) In a time dependent background, if chemical equilibrium exists, there will be another effect which could make $\mu_A+\mu_B+\ldots\ne\mu_C+\mu_D+\ldots$.\\
Recall that the $\delta^4(p_A+p_B\ldots -p_C-p_D\ldots)$ in the Boltzmann equation comes from energy-momentum conservation. Especially, the energy conservation could be derived by time translational invariance of the Lagrangian. If the Lagrangian depends on time explicitly,\footnote{In our case, the fundamental Lagrangian does not depend on time explicitly. When the axion/Majoron is treated as time dependent classical background, the effective Lagrangian depends on time explicitly.} then the time translational invariance no longer exists and the delta function may need to be modified.\\
This nonconservation of energy may be seen by Noether's theorem. The energy-momentum tensor for a time dependent Lagrangian satisfies
\begin{equation}
\partial_a T^a_{\,\,\,b}=-\frac{\partial\mathcal{L}}{\partial x^b}=-\delta^0_b\frac{\partial\mathcal{L}}{\partial t}
\end{equation}
More details of  Noether's theorem for a time dependent Lagrangian may be found in Appendix \ref{Noether}.\\
Consider the following modification of the Boltzmann equation
\begin{eqnarray}
\frac{dn_A}{dt}&\equiv&-\int  d\Pi_Ad\Pi_B\ldots d\Pi_C d\Pi_D\ldots\nonumber\\
&&\times (2\pi)^4\times\delta(E_A+E_B+\cdots-E_C-E_D\cdots-\Delta)\times\delta^3(\vec{p}_A+\vec{p}_B\ldots -\vec{p}_C-\vec{p}_D\ldots)\nonumber\\
&&\times \left[\begin{array}{l}\,\,\,\,\vert\mathcal{M}\vert^2_{A+B+\cdots\to C+D+\cdots}f_A f_B\cdots (1\pm f_C)(1\pm f_D)\cdots\\-\vert\mathcal{M}\vert^2_{C+D+\cdots\to A+B+\cdots} f_Cf_D\cdots (1\pm f_A)(1\pm f_B)\cdots\end{array}\right]\label{1}
\end{eqnarray}
If we still have
\begin{equation}
\vert\mathcal{M}\vert^2_{A+B+\ldots\to C+D+\ldots}=\vert\mathcal{M}\vert^2_{C+D+\ldots\to A+B+\ldots}\label{2}
\end{equation}
then in chemical equilibrium one could derive
\begin{equation}
\mu_A+\mu_B+\cdots=\mu_C+\mu_D+\cdots+\Delta \label{3}
\end{equation}
\subsubsection{The equilibrium from the viewpoint of basis (B)}\label{B}
In  basis (B), everything is time independent, and all terms are $T$ invariant (see Appendix \ref{8} for why $\delta_l l^\dagger l$ and $\delta_q q_i^\dagger q_i$ are $T$ invariant). Therefore, in basis (B) we could use Eq. (\ref{flat2}) in chemical equilibrium.\\
The fermions in the toy model participate in two types of interactions, namely the perturbative gauge interaction and the nonperturbative anomalous gauge interaction.\footnote{The anomalous gauge interaction may be an instanton or sphaleron  interaction.}\\
(1) When the perturbative gauge interaction is in equilibrium with a unbroken gauge symmetry   we  have
\begin{equation}
 \mu_W=0.
\end{equation}
This allows us to define the chemical potential  $\mu_l$ and $\mu_{q_i}$ for each doublet. Also, we have $\mu_{\bar{l}}=-\mu_{{l}}$ and $\mu_{\bar{q}_i}=-\mu_{{q}_i}$, similarly for the effective chemical potential $\bar{\mu}_{\bar{l}}=-\bar{\mu}_{{l}}$ and $\bar{\mu}_{\bar{q}_i}=-\bar{\mu}_{{q}_i}$. With this in mind, we will not repeatedly write down the chemical potentials or effective chemical potentials for antiparticles.\\
(2) When, in addition, the anomalous gauge interaction is in equilibrium
\begin{equation}
\mu_l+\sum_{i=1}^3\mu_{q_i}=0\qquad\Longrightarrow\qquad \bar{\mu}_l+\sum_{i=1}^3\bar{\mu}_{q_i}=-\delta \label{equili}
\end{equation}
Note that while $\mu_l+\sum_{i=1}^3\mu_{q_i}=0$ we may have $\bar{\mu}_l+\sum_{i=1}^3\bar{\mu}_{q_i}\ne0$. Also  $\delta$ only enters the equation of the effective chemical potential for the anomalous interaction but not the perturbative gauge interaction, and the energy shifts $\delta_l$ and $\delta_q$ do not come separately in the effective chemical potential equations, only the combination $\delta=\delta_l+3\delta_q$ matters. The above facts can be explained more easily by changing basis.
\subsubsection{The equilibrium from the viewpoint of basis (A)}\label{A}
We can solve the entire problem in basis (B), but as physics is independent of  basis, it worthwhile to share the viewpoint of basis (A).\\
As is argued earlier in Sec. \ref{invariant}, the particle number density $n$, the occupancy $f(\vec{p})$ and the effective chemical potential $\bar{\mu}$ are invariant quantities under the basis changes we consider. In basis (A), the amount of energy shift is zero, so
\begin{equation}
\mu_l=\bar{\mu}_l\qquad\qquad \mu_{q_i}=\bar{\mu}_{q_i}
\end{equation}
With the result obtained from basis (B) in Eq. (\ref{equili}), and the invariance of effective chemical potential, we find in basis (A)
\begin{equation}
\bar{\mu}_l+\sum_{i=1}^3\bar{\mu}_{q_i}=-\delta\qquad\Longrightarrow\qquad \mu_l+\sum_{i=1}^3\mu_{q_i}=-\delta
\end{equation}
A more careful comparison of different  basis (at the level of Boltzmann equation) in the following section \ref{Boltzmann} shows that the reason to have $\mu_l+\sum_{i=1}^3\mu_{q_i}\ne0$  in basis (A) is that the time dependent term $\theta(x)W\tilde{W}$ causes the situation described in Eqs.  (\ref{1}), (\ref{2}), (\ref{3}). Furthermore, from the viewpoint of basis (A), there is no surprise that $\delta$ only affects the effective chemical potential equation of the anomalous interactions (but not the perturbative gauge interaction) since it is in $\theta W\tilde{W}$.
\subsubsection{Compare basis (A) and (B) at the level of Boltzmann equation}\label{Boltzmann}
The last two sections \ref{B} and \ref{A} mainly focus on equilibrium. Some more information can be seen at the level of the Boltzmann equations, and it makes the correspondence between the two bases more clear.\\
Consider the following anomalous interaction:
\begin{equation}
l+q_1\rightleftharpoons \bar{q}_2+\bar{q}_3
\end{equation}
Let the partial rate\footnote{It is called partial rate because the number $l$ particle can be changed by other interactions. } for the change of the number density for $l$ particles, due to this interaction, be
\begin{equation}
\bigg(\frac{dn_l}{dt}\bigg)_{l+q_1\rightleftharpoons \bar{q}_2+\bar{q}_3}
\end{equation}
In basis (B), as everything in the Lagrangian is time independent, and all terms are $T$ invariant, the Boltzmann equation for this process is
\begin{eqnarray}
(B)\qquad\qquad\bigg(\frac{dn_l}{dt}\bigg)_{l+q_1\rightleftharpoons \bar{q}_2+\bar{q}_3}&=&-\int d\Pi_l d\Pi_{q_1} d\Pi_{\bar{q}_2} d\Pi_{\bar{q}_3}
\times (2\pi)^4\delta^4(p_l+p_{q_1}-p_{\bar{q}_2}-p_{\bar{q}_3})\qquad\qquad\nonumber\\
&&\times \left[\begin{array}{l}\,\,\,\,\vert\mathcal{M}\vert^2_{l+q_1\to \bar{q}_2+\bar{q}_3}\,\,f_l f_{q_1} (1- f_{\bar{q}_2})(1- f_{\bar{q}_3})\\-\vert\mathcal{M}\vert^2_{\bar{q}_2+\bar{q}_3\to l+q_1}\,\, f_{\bar{q}_2}f_{\bar{q}_3} (1- f_{l})(1- f_{q_1})\end{array}\right] \label{Bol B}
\end{eqnarray}
the energy is conserved, and  from $T$ invariance, we have
\begin{equation}
(B)\qquad\qquad\qquad\qquad\vert\mathcal{M}\vert^2_{l+q_1\to \bar{q}_2+\bar{q}_3}=\vert\mathcal{M}\vert^2_{\bar{q}_2+\bar{q}_3\to l+q_1} \qquad\qquad\qquad\qquad
\end{equation}
As argued in Sec. \ref{invariant}, when going from basis (B) to basis (A), the number density $n_l$, the occupancy $f_l(\vec{p})$ $f_{q_1}(\vec{p})$ $f_{\bar{q}_2}(\vec{p})$ $f_{\bar{q}_3}(\vec{p})$ are invariant quantities, and the fermion energy changes according to the energy shifts. The only consistent way is to have the following Boltzmann equation in basis (A):
\begin{eqnarray}
(A)\qquad\qquad\bigg(\frac{dn_l}{dt}\bigg)_{l+q_1\rightleftharpoons \bar{q}_2+\bar{q}_3}&=&-\int d\Pi_l d\Pi_{q_1} d\Pi_{\bar{q}_2} d\Pi_{\bar{q}_3}\nonumber\\
&&\times (2\pi)^4\delta(E_l+E_{q_1}-E_{\bar{q}_2}-E_{\bar{q}_3}+\delta)\times \delta^3(\vec{p}_l+\vec{p}_{q_1}-\vec{p}_{\bar{q}_2}-\vec{p}_{\bar{q}_3})\qquad\nonumber\\
&&\times \left[\begin{array}{l}\,\,\,\,\vert\mathcal{M}\vert^2_{l+q_1\to \bar{q}_2+\bar{q}_3}\,\,f_l f_{q_1} (1- f_{\bar{q}_2})(1- f_{\bar{q}_3})\\-\vert\mathcal{M}\vert^2_{\bar{q}_2+\bar{q}_3\to l+q_1}\,\, f_{\bar{q}_2}f_{\bar{q}_3} (1- f_{l})(1- f_{q_1})\end{array}\right]
\end{eqnarray}
with
\begin{equation}
(A)\qquad\qquad\qquad\qquad\vert\mathcal{M}\vert^2_{l+q_1\to \bar{q}_2+\bar{q}_3}=\vert\mathcal{M}\vert^2_{\bar{q}_2+\bar{q}_3\to l+q_1} \qquad\qquad\qquad\qquad
\end{equation}
From this Boltzmann equation one can directly derive the relation
\begin{equation}
\mu_l+\sum_{i=1}^3\mu_{q_i}=-\delta .
\end{equation}
This shows that the reason $\mu_l+\sum_{i=1}^3\mu_{q_i}\ne0$  in basis (A) is that the time dependent term $\theta(x)W\tilde{W}$ causes the situation described in Eqs. (\ref{1}), (\ref{2}), (\ref{3}). In other words, from the viewpoint of basis (A),  the effect of the operator $\theta(x)W\tilde{W}$ in the anomalous interaction  ${l+q_1\rightleftharpoons \bar{q}_2+\bar{q}_3}$ is to make the sum of the energies of the incoming particles not equal to the sum of the energies of the outgoing particles, i.e.
\begin{equation}
E_{\bar{q}_2}+E_{\bar{q}_3}=E_{l}+E_{q_1}+\delta \label{delta} .
\end{equation}
An independent proof of Eq. (\ref{delta}), using Noether's theorem, can be found in Appendix \ref{Noether}.
\subsection{Insight from different bases}
Here,  we remark that  it is very useful to go back and forth between different bases, and there are insights easier seen in one basis rather than another. For the toy models described above:

(1) In basis (A), the modification of the dispersion relation of the gauge boson, Eq. (\ref{dispersion}), can be derived at the classical level, while in basis (B), it can be seen only after doing a 1-loop calculation. Therefore, the modification of the dispersion relation is best seen in basis (A). Furthermore, by the invariance of physics, and change of basis, we predict what the 1-loop diagram should give us before doing any calculation. This shows the power of basis changes.

(2) In basis (B), the Lagrangian is time independent, and all terms in the Lagrangian are T invariant. Therefore, the Boltzmann equation looks most familiar,  see Eq. (\ref{Bol B}), and the  chemical potential equations  and  effective chemical potential equations are most easily derived. By changing basis rather than direct calculation we find that the $\theta(x)W\tilde{W}$ term, from the viewpoint of basis (A), is responsible for the energy nonconservation in the anomalous interactions, see Eq. (\ref{delta}). [A direction calculation in basis (A) using Noether's theorem which confirms the result is provided in Appendix \ref{Noether}.]

(3) In basis (B), it is not straightforward to see why $\delta_l$ and $\delta_q$ do not separately enter the effective chemical potential equations, and only the combination $\delta=\delta_l+3\delta_q$ matters. Nevertheless, by changing basis, and the invariance of physics, one can argue that only $\delta$ matters by choosing a basis, for example, with $\delta_l=\delta$ and $\delta_q=0$.\footnote{The basis with with $\delta_l=\delta$ and $\delta_q=0$ is a special case of basis (B).}

\section{The chemical equilibrium equations for axion/Majoron leptogenesis models}\label{realistic}
In this section, we work out the equations for the effective chemical potentials in realistic models. Consider the SM Lagrangian with neutrino mass and added in energy shifts:
\begin{equation}
\mathcal{L}=\mathcal{L}_{kinetic}+\mathcal{L}_{Yukawa}+\mathcal{L}_{gauge}+\mathcal{L}_{Higgs}+\mathcal{L}_M+\mathcal{L}_\delta \label{5} .
\end{equation}
In the following we only explicitly write down the Lagrangian for one family of quarks and leptons while keeping in mind there could be $N_f$ families (we are mostly interested in $N_f=3$). We assume there is one Higgs doublet. We neglect the $CP$-violating mixing in the Yukawa couplings for this problem, and we will assume the chemical potentials are independent of family.
\begin{eqnarray}
\mathcal{L}_{kinetic}&=& l^\dagger i\bar{\sigma}^a D_a l+ \bar{e}^\dagger i\bar{\sigma}^a D_a \bar{e}+q^\dagger i\bar{\sigma}^a D_a q+\bar{u}^\dagger i\bar{\sigma}^a D_a \bar{u}+\bar{d}^\dagger i\bar{\sigma}^a D_a \bar{d}     \\
\mathcal{L}_{Yukawa}&=& -g_e(\bar{e} H^\dagger l-l^\dagger H\bar{e}^\ast)
-g_d(\bar{d}H^\dagger q-q^\dagger H\bar{d}^\ast)-g_u(\bar{u}\tilde{H}^\dagger q - q^\dagger \tilde{H}\bar{u}^\ast)        \\
\mathcal{L}_{gauge}&=&   -\frac{1}{2g_s^2}tr(G_{ab}G^{ab})   -\frac{1}{2g^2}tr(W_{ab}W^{ab})- \frac{1}{4 g'^2}B_{ab}B^{ab}     \\
\mathcal{L}_{Higgs}&=& (D_a \phi)^\dagger(D^a \phi)-V(\phi)\\
\mathcal{L}_M&=&\frac{g^2_\nu}{2M}[(\tilde{H}^\dagger l)(\tilde{H}^\dagger l)-(l^\dagger\tilde{H})(l^\dagger\tilde{H}) ]                                      \\
\mathcal{L}_\delta&=& -\delta_q J^0_Q-\delta_l J^0_L
\end{eqnarray}
where the two component spinor indices are antisymmetrized (see Appendix \ref{ssec:spinors}).
We assume there are heavy neutrinos, and we have integrated them out because we are interested in the physics at a much lower energy scale than the heavy neutrino mass $M$. The term $\mathcal{L}_M$ is the dimension-5 Weinberg operator that is obtained by integrating out the heavy neutrinos. Here $M$ is real, and $J^a_Q$ and $J^a_L$ are the quark and lepton currents (the baryon current  $J^a_B\equiv\frac{1}{3}J^a_Q$)
\begin{eqnarray}
J^a_L&=&l^\dagger\bar{\sigma}^al-\bar{e}^\ast \bar{\sigma}^a\bar{e}
\\
J^a_Q&=&q^\dagger\bar{\sigma}^aq-\bar{u}^\ast \bar{\sigma}^a\bar{u}-\bar{d}^\ast \bar{\sigma}^a\bar{d}
\end{eqnarray}
For later convenience let us define $\delta\equiv 3\delta_q+\delta_l$, and we will use $\delta$ and $\delta_l$ as two independent variables (instead of using $\delta_q$ and $\delta_l$). We assume in this problem $\delta$ and $\delta_l$  change with time slowly enough and in chemical equilibrium we can treat them as constants. Also, we assume $\delta\ll T$ and $\delta_l\ll T$, where $T$ is the temperature of the thermal plasma.
\subsection{Energy shifts  from axion/Majoron leptogenesis models}
In this section, we show that the axion leptogenesis model in \cite{Kusenko:2014uta} is equivalent to our Lagrangian (\ref{5}) with $\delta\ne0$ and $\delta_l=0$, and the Majoron leptogenesis model in \cite{Ibe:2015nfa} is equivalent to our Lagrangian (\ref{5}) with $\delta=0$ and $\delta_l\ne0$. In order to verify this statement, a basis change is needed and the change of Lagrangian due to the change in the path integral measure is taken into account.
\subsubsection{For the  axion leptogenesis model in \cite{Kusenko:2014uta}}
In our notation, the Lagrangian for the axion leptogenesis model in \cite{Kusenko:2014uta} looks like
\begin{equation}
\mathcal{L}=\mathcal{L}_{kinetic}+\mathcal{L}_{Yukawa}+\mathcal{L}_{gauge}+\mathcal{L}_{Higgs}+\mathcal{L}_M+\mathcal{L}_{anomaly}\label{L original}
\end{equation}
where
\begin{equation}
\mathcal{L}_{anomaly}=-\frac{\theta(x)}{16\pi^2}[tr(W_{ab}\tilde{W}^{ab})-2B_{ab}\tilde{B}^{ab}]\qquad\qquad \theta(x)=\frac{a(x)}{f_a}
\end{equation}
Here $a(x)$ is the electroweak axion field which we treat as a classical background, and $f_a$ is the axion decay constant. We consider, as in \cite{Kusenko:2014uta}, a homogeneous and time dependent background, so we have
\begin{equation}
\partial_b{\theta}=\frac{\partial_b a}{f_a}\equiv(\Delta,0,0,0) .
\end{equation}
With the following basis change (vector rotations on quarks)\footnote{We do not rotate the leptons, since this would induce phases in the dim-5 Weinberg operator.}
\begin{eqnarray}
q&\to& e^{i\theta_2(x)}q \nn\\
\bar{u}&\to& e^{-i\theta_2(x)}\bar{u}\\
\bar{d}&\to& e^{-i\theta_2(x)}\bar{d} \nn
\end{eqnarray}
and
\begin{equation}
\theta_2(x)=\frac{\theta(x)}{3N_f}
\end{equation}
(refer to Appendix \ref{Fujikawa} for how the Lagrangian changes), we find the Lagrangian in this new basis is just described as Eq. (\ref{5}) with
\begin{equation}
 \delta=\frac{\Delta}{N_f}=\frac{\dot{a}}{N_f f_a}\qquad\qquad\delta_l=0 .
\end{equation}
In the relevant temperature range, $\delta\ll T$ is satisfied.
\subsubsection{For the Majoron leptogenesis model in \cite{Ibe:2015nfa}}
In our notation, the Lagrangian for the Majoron leptogenesis model in \cite{Ibe:2015nfa} looks like
\begin{equation}
\mathcal{L}=\mathcal{L}_{kinetic}+\mathcal{L}_{Yukawa}+\mathcal{L}_{gauge}+\mathcal{L}_{Higgs}+\mathcal{L}'_{M}
\end{equation}
where
\begin{equation}
\mathcal{L}'_{M}=\frac{g^2_\nu}{2M}[e^{-i\theta(x)}(\tilde{H}^\dagger l)(\tilde{H}^\dagger l)-e^{i\theta(x)}(l^\dagger\tilde{H})(l^\dagger\tilde{H}) ]
\end{equation}
here compare with the notation in \cite{Ibe:2015nfa}
\begin{equation}
\theta(x)=\frac{\chi(x)}{\sqrt{2}v_{B-L}}\qquad\qquad  \partial_a\theta=\frac{\partial_a\chi}{\sqrt{2}v_{B-L}}\equiv(\Delta_l,0,0,0)
\end{equation}
where $\chi(x)$ is the Majoron field which we treated as a classical background, and $v_{B-L}$ is the $B-L$ breaking scale which is assumed to be roughly the same scale as the heavy neutrino mass $M$.
With the following basis change (vector rotation on quarks and leptons):
\begin{eqnarray}
l&\to& e^{i\theta_1(x)}l\nn\\
\bar{e}&\to& e^{-i\theta_1(x)}\bar{e} \nn \\
q&\to& e^{i\theta_2(x)}q \\
\bar{u}&\to& e^{-i\theta_2(x)}\bar{u} \nn \\
\bar{d}&\to& e^{-i\theta_2(x)}\bar{d} \nn
\end{eqnarray}
and
\begin{equation}
\theta_1(x)=\frac{1}{2}\theta(x)\qquad\qquad \theta_2(x)=-\frac{1}{6}\theta(x)\qquad\Longrightarrow\qquad \theta_1+3\theta_2=0
\end{equation}
(refer to Appendix \ref{Fujikawa} for how the Lagrangian changes), we find the Lagrangian in this new basis is just described as Eq. (\ref{5}) with
\begin{equation}
 \delta=0\qquad\qquad\delta_l=\frac{\Delta_l}{2}=\frac{\dot{\chi}}{2\sqrt{2}v_{B-L}}   \label{eq:deltal}
\end{equation}
In the relevant temperature range, $\delta_l\ll T$ is satisfied.   Note, the second term in Eq. (\ref{eq:deltal}), i.e. $\delta_l$,  is the same value found in Ref. \cite{Ibe:2015nfa}.  However the derivation
in \cite{Ibe:2015nfa} did not take into account Fujikawa's result for the change in the fermion path integral measure under the basis change and the fact that $\delta = 0$.

\subsection{Effective chemical potential in the early universe}\label{4}
We work out the equations for the effective chemical potentials when the relevant process is in chemical equilibrium. We will use notation very similar to that in Ref. \cite{Buchmuller:2005eh}, and we consider the following result to be the generalization of the result in \cite{Buchmuller:2005eh} for the types of slowly changing time dependent background fields described above. The chemical potential will be used in intermediate steps , but we would like the final result to be written in terms of effective chemical potentials because the effective chemical potential is invariant under basis changes. The intermediate steps with chemical potential will depend on the specific  basis chosen, while the final result in terms of the effective chemical potential is independent of basis.\footnote{We  use the effective chemical potential for fermions but not for gauge bosons or Higgs because in our problem, gauge bosons and Higgs never get energy shifts, i.e. neglecting possible small energy shifts for the gauge bosons.}\\
a) In the early universe before electroweak symmetry breaking, the chemical potentials of the gauge bosons vanish
\begin{equation}
\mu_B=\mu_W=\mu_g=0 .
\end{equation}
b) When fermion and Higgs interactions with gauge bosons are in equilibrium, it is possible to assign  a single chemical potential for each fermion or Higgs multiplet. The chemical potential of particles and antiparticles add up to zero. Moreover, $\delta$ and $\delta_l$ shifts for particles and antiparticles are opposite. Therefore, the effective chemical potential for particles and antiparticles adds up to zero. With this in mind, we only write down the chemical potential for particles (not antiparticles). We also assume the chemical potentials are independent of family and therefore we drop the family indices.
Given the following chemical potentials and effective chemical potentials we find relations among them when interaction rates are in equilibrium, i.e. they are fast compared to the Hubble expansion rate.  We have
\begin{equation}
\mu_l\qquad\mu_e\qquad\mu_q\qquad\mu_u\qquad\mu_d\qquad\mu_H \qquad(\textrm{chemical potentials})
\end{equation}
\begin{equation}
\quad\,\,\qquad\bar{\mu}_l\qquad\bar{\mu}_e\qquad\bar{\mu}_q\qquad\bar{\mu}_u\qquad\bar{\mu}_d\qquad\bar{\mu}_H\qquad(\textrm{effective chemical potentials}) .
\end{equation}
Since $\bar{\mu}_H=\mu_H$ in any basis, we will use $\mu_H$ for both chemical potential and effective chemical potential of the Higgs doublet.\\
c) When the QCD anomaly is in  equilibrium, we have
\begin{equation}
2\mu_q=\mu_u+\mu_d\qquad\Longrightarrow\qquad 2\bar{\mu}_q=\bar{\mu}_u+\bar{\mu}_d .
\end{equation}
In the derivation we used the fact that all quarks shift by the same amount $\delta_q$.\\
d) When the $SU(2)_L$ anomaly is in equilibrium, we have
\begin{equation}
3\mu_q+\mu_l=0\qquad\Longrightarrow\qquad 3\bar{\mu}_q+\bar{\mu}_l=-(3\delta_q+\delta_l)=-\delta .  \label{eq:anomaly}
\end{equation}
e) When Yukawa coupling interactions are in equilibrium, we have
\begin{eqnarray}
\mu_q&=&\mu_d+\mu_H\qquad\Longrightarrow\qquad\bar{\mu}_q=\bar{\mu}_d+\mu_H\\
\mu_q&=&\mu_u-\mu_H\qquad\Longrightarrow\qquad\bar{\mu}_q=\bar{\mu}_u-\mu_H\\
\mu_l&=&\mu_e+\mu_H\qquad\Longrightarrow\qquad\bar{\mu}_l=\bar{\mu}_e+\mu_H .
\end{eqnarray}
f) The requirement of a hypercharge neutral universe, $\sum Y=0$, constrains $\bar{\mu}_i$ directly, rather than $\mu_i$ since what is relevant is the number density
\begin{equation}
\bar{\mu}_q+2\bar{\mu}_u-\bar{\mu}_d-\bar{\mu}_l-\bar{\mu}_e+\frac{2}{N_f}\mu_H=0 .
\end{equation}
Here we have used the  approximation that for $\bar{\mu}_i\ll T$, each internal degree of freedom gives you
\begin{eqnarray}
(Fermion)&&\qquad n_i-n_{\bar{i}}\simeq\frac{1}{6}\bar{\mu}_iT^2 \nn\\
(Boson)&&\qquad n_i-n_{\bar{i}}\simeq\frac{1}{3}\bar{\mu}_iT^2 . \nn
\end{eqnarray}
g) The lepton number changing $\Delta L=2$ interaction gives
\begin{equation}
\Delta L=2\qquad\qquad \bar{H}+\bar{H}\rightleftharpoons l+l\qquad\qquad \bar{l}+\bar{H}\rightleftharpoons l+H .
\end{equation}
When it is in equilibrium, and notice the Lagrangian (\ref{5}) is $T$ invariant and time independent, we have
\begin{equation}
\mu_l+\mu_H=0\qquad\Longrightarrow\qquad\bar{\mu}_l+\mu_H=-\delta_l .
\end{equation}
We observe that the $\delta$, which could come from a time dependent electroweak axion background, only appears in the effective chemical potential equation for the electroweak anomaly, and $\delta_l$ which could come from a time dependent Majoron background only appears in the effective chemical potential equation for  $\Delta L=2$ interactions.

\subsection{Phenomenological implications }
\subsubsection{The equilibrium point in the limit $SU(2)_L$ sphaleron is turned off}
In this section, the phrase turned off means the theoretical limit in which some specific interaction rate goes to zero. It should be understood as a theoretical limit which is useful to obtain some insight. This limit is not necessarily realized in realistic situations.  However it is a good approximation when the electroweak sphaleron rate satisfies, $\Gamma_{B+L} \ll H$.   This is in fact relevant to the axionic leptogenesis model \cite{Kusenko:2014uta} for $T > 10^{13}$ GeV.

We are trying to solve for the baryon number density $n_B$ and lepton number density $n_L$. In the early universe, the $SU(2)_L$ anomalous interaction is the sphaleron interaction, and if it is turned off, we cannot use (d). Let us solve the equilibrium effective chemical potentials when all other interactions are in equilibrium.
The useful relations of fermion effective chemical potential from (a), (b), (c), (e), (f), (g) are
\begin{eqnarray}
(\textrm{Yukawa})&&\bar{\mu}_q=\bar{\mu}_d+\mu_H\\
(\textrm{Yukawa})&&\bar{\mu}_q=\bar{\mu}_u-\mu_H\\
(\textrm{Yukawa})&&\bar{\mu}_l=\bar{\mu}_e+\mu_H\\
(Y=0)&&\bar{\mu}_q+2\bar{\mu}_u-\bar{\mu}_d-\bar{\mu}_l-\bar{\mu}_e+\frac{2}{N_f}\mu_H=0\\
(L)&&\bar{\mu}_l+\mu_H=-\delta_l .
\end{eqnarray}
Solving the equilibrium in terms of  $\delta_l$ and $\bar{\mu}_q$ we find ($\delta$ does not enter this result)
\begin{eqnarray}
&&\bar{\mu}_l=-\frac{2N_f+1}{3N_f+1}\delta_l+\frac{N_f}{3N_f+1}\bar{\mu}_q  \label{eq:mubar} \nn\\
&&\mu_H=-\frac{N_f}{3N_f+1}(\delta_l+\bar{\mu}_q)          \nn   \\
&&\bar{\mu}_e=-\frac{N_f+1}{3N_f+1}\delta_l+\frac{2N_f}{3N_f+1}\bar{\mu}_q            \\
&&\bar{\mu}_d=\frac{N_f}{3N_f+1}\delta_l+\frac{4N_f+1}{3N_f+1}\bar{\mu}_q    \nn         \\
&&\bar{\mu}_u=-\frac{N_f}{3N_f+1}\delta_l+\frac{2N_f+1}{3N_f+1}\bar{\mu}_q  \nn
\end{eqnarray}
and
\begin{eqnarray}
&&B=4N_f\bar{\mu}_q\\
&&L=N_f\bigg(-\frac{5N_f+3}{3N_f+1}\delta_l+\frac{4N_f}{3N_f+1}\bar{\mu}_q\bigg)\\
&&B-L=N_f\bigg( \frac{5N_f+3}{3N_f+1}\delta_l+\frac{8N_f+4}{3N_f+1}\bar{\mu}_q   \bigg)
\end{eqnarray}
where $B$ and $L$ are defined such that the net baryon number density $n_B$ and the net lepton number density $n_L$ can be written as
\begin{eqnarray}
n_B&\equiv& n_b-n_{\bar{b}}\simeq \frac{BT^2}{6} \qquad\Longrightarrow\qquad B=N_f(2\bar{\mu}_q+\bar{\mu}_u+\bar{\mu}_d)  \\
 n_L&\equiv& n_l-n_{\bar{l}}\simeq\frac{LT^2}{6}  \qquad\Longrightarrow\qquad L=N_f(2\bar{\mu}_l+\bar{\mu}_e) .
\end{eqnarray}
As we are interested in the early universe when the lepton number is generated. In the limit the $SU(2)_L$ sphaleron is turned off, $n_B$ does not change at that period of time. We are interested in the initial condition $n_B=0$, which gives you
\begin{equation}
B=0\qquad\Longrightarrow\qquad 2\bar{\mu}_q+\bar{\mu}_u+\bar{\mu}_d=0\qquad\Longrightarrow\qquad \bar{\mu}_q=0 .  \label{eq:B}
\end{equation}
Plugging the result of Eq. (\ref{eq:B}) into Eq. (\ref{eq:mubar}) we find\footnote{In this case, only $\delta_l$ matters and it is reasonable.}
\begin{eqnarray}
&&\bar{\mu}_q=0 \nn\\
&&\bar{\mu}_l=-\frac{2N_f+1}{3N_f+1}\delta_l \nn\\
&&\mu_H=-\frac{N_f}{3N_f+1}\delta_l \\
&&\bar{\mu}_e=-\frac{N_f+1}{3N_f+1}\delta_l \nn\\
&&\bar{\mu}_d=\frac{N_f}{3N_f+1}\delta_l \nn\\
&&\bar{\mu}_u=-\frac{N_f}{3N_f+1}\delta_l \nn
\end{eqnarray}

In the model described in \cite{Kusenko:2014uta}, we argued that $\delta_l=0$.  We find the equilibrium value of the effective chemical potentials ($\bar{\mu}_q$ $\bar{\mu}_l$ $\bar{\mu}_e$ $\bar{\mu}_d$ $\bar{\mu}_u$  $\mu_H$) to be zero. Therefore, $B=L=0$ and no asymmetry could be generated in the limit the $SU(2)_L$ sphaleron interaction is turned off.\footnote{From the viewpoint of the original basis Eq. (\ref{L original}), the effect of the time dependent axion background is to make energy not conserved in sphaleron interaction, and when the sphaleron is turned off, it could not affect the equilibrium and therefore, could not produce asymmetry. For another possible basis in which the anomaly term is canceled by a vector rotation on leptons, time dependence in the dim-5 Weinberg operator and the energy shifts on leptons both have effect, and in the end the same result is obtained.}  This is a different result than obtained in \cite{Kusenko:2014uta}, in which only the $\Delta L=2$ interaction rate, $\Gamma_L$, enters the Boltzmann equations and $\delta_l$ was assumed to be nonvanishing.  Note, if the sphaleron interaction is not completely turned off, i.e. we do not neglect the results of Eq. (\ref{eq:anomaly}),  then the equilibrium value of $\bar{\mu}_q \neq 0$ and a baryon asymmetry will be generated by $\delta$.

This result suggests that for the model in \cite{Kusenko:2014uta}, nonzero $B$ must be generated at the time nonzero $B-L$ is generated, otherwise $B-L=0$. The sphaleron interaction rate per particle satisfies $\Gamma_{B+L} < H$ when $T>10^{12}$ GeV (since the sphaleron decouples at $T>10^{12}$ GeV, see for example \cite{Buchmuller:2005eh}) and according to the data in \cite{Kusenko:2014uta} the $\Delta L=2$ interaction rate $\Gamma_L>H$ for $T > 10^{13}$ GeV. Therefore, at $T>10^{13}$ GeV we expect the amount of $B-L$ generated is controlled by the smallness of $\Gamma_{B+L}$ rather than $\Gamma_{L}$, since $\Gamma_L > H \gg \Gamma_{B+L}$. We expect the $B-L$ generated at $T>10^{13}$ GeV to be less efficient than described in \cite{Kusenko:2014uta}.
\subsubsection{A subleading order effect }\label{sub-leading observable}
We notice a subleading order effect: the modification of the dispersion relation of the gauge boson, [similar to Eq. (\ref{dispersion})].
\begin{eqnarray}
(W^a \,\,\,\,\textrm{field})\qquad\qquad\omega^2&=&k^2\pm \frac{g^2}{8\pi^2}k(N_f\delta)\\
(B^a \,\,\,\,\textrm{field})\qquad\qquad\omega^2&=&k^2\mp \frac{g'^2}{8\pi^2}k(N_f\delta) .
\end{eqnarray}
This effect exists in the axionic leptogenesis model \cite{Kusenko:2014uta} but does not exist in the Majoron model \cite{Ibe:2015nfa}.
It is an effect at subleading order, has similar $g^2$ suppression  as the thermal correction to the gauge boson dispersion relation, and when calculating the effective chemical potentials we neglected this effect.

The axion modification of the dispersion relation is different from a thermal correction:\\
(1) It is a zero temperature effect.\\
(2) The correction is linear in $k$ and therefore it is not a mass term. On the other hand, the thermal effect is a mass term ($m\sim gT$ for $k\gg gT$).\\
(3) The thermal correction treats the two circularly polarized modes in the same way, while the axion correction treats $\pm$ circularly polarized modes (or helicity) differently.\\

It may be interesting to investigate whether this modification of gauge boson dispersion relations leads to any observable.
A possible observable due to a cold axion background modified dispersion relation of a photon is discussed in \cite{Espriu:2014lma}.  The dispersion relation discussed is of very similar origin  as the one we consider, but the energy scale is very different, and it could make the observable (if it exists) very different.

The fact that an axion background treats the two circularly polarized modes ($\pm$ helicity states) of gauge bosons differently could result in a nonzero helicity density $\mathcal{H}\ne 0$. The definition of helicity density $\mathcal{H}$  in the context of electrodynamics may be found in \cite{Vachaspati:2001nb} together with its possible origin during an electroweak phase transition ($T\sim 100$ GeV). Our result indicates that  the axionic leptogenesis model \cite{Kusenko:2014uta}  could give rise to a $\mathcal{H}\ne 0$ (for the gauge fields $B^a$ and $W^a$) at a much higher temperature scale ($T\sim 10^{12}$ GeV). It is not clear to us whether such a $\mathcal{H}\ne 0$ in the early universe could induce an observable effect today.

\section{Acknowledgements}
One of the authors, B.S., would like to thank B. Charles Bryant for helpful discussion about Majorana neutrinos,  thank Hyung Do Kim for a discussion about electroweak vacuum angle and thank Mingzhe Li from whose talk B.S. learned for the first time that the Chern-Simons term could modify the dispersion relation of photons.   S.R.~received partial support for this work from DOE/ DE-SC0011726.   During the final stages of this paper, S.R. was supported by the Munich Institute for Astro- and Particle Physics (MIAPP) of the DFG cluster of excellence ``Origin and Structure of the Universe."

\appendix
\section{Some more details}
\subsection{Our notation of gauge fields  and spinors}
\label{ssec:spinors}
For spinors, we only use left-handed Weyl spinors.
Here is a comparison between different notations
\begin{equation}
\psi\equiv\left(  \begin{array}{c} \psi_L\\ \psi_R    \end{array}      \right)=\left(  \begin{array}{c} \alpha\\ i\sigma^2\beta ^\ast  \end{array}      \right)\qquad\qquad\sigma^a\equiv(1,\sigma^i)\qquad\qquad \bar{\sigma}^a\equiv(1,-\sigma^i)
\end{equation}
\begin{eqnarray}
\mathcal{L}=\bar{\psi}(i\slashed{\partial}-m)\psi&=&\psi^\dagger_L i\bar{\sigma}^a\partial_a \psi_L +\psi^\dagger_R i{\sigma}^a\partial_a \psi_R-m(\psi^\dagger_L\psi_R+\psi_R^\dagger\psi_L)\nonumber\\
&=& \alpha^\dagger i\bar{\sigma}^a\partial_a \alpha + \beta^\dagger i\bar{\sigma}^a\partial_a \beta + m [(\alpha \beta) - (\alpha^\ast \beta^\ast)]
\end{eqnarray}  where $(\alpha \beta) \equiv \alpha^T (i \sigma_2) \beta$.
Both $\alpha$ and $\beta$ are left-handed Weyl spinors.
\begin{equation}
\bar{\psi} i\gamma^a A_a \psi=\psi_L^\dagger i\bar{\sigma}^a A_a\psi_L+\psi_R^\dagger i\sigma^a A_a \psi_R=\alpha^\dagger i\bar{\sigma}^a A_a\alpha + \beta^\dagger i\bar{\sigma}^a (-A_a^T)\beta
\end{equation}
As $-A_a^T$ is  $A_a$ in the conjugate representation, when the spinor switches $\psi_R\to \beta$, the representation switches into its conjugate representation.\\
Let us look at how the vector  current looks like in different notations
\begin{equation}
j^a\equiv \bar{\psi}\gamma^a\psi=\psi^\dagger_L \bar{\sigma}^a \psi_L + \psi_R^\dagger \sigma^a \psi_R=\alpha^\dagger \bar{\sigma}^a \alpha -\beta^\dagger \bar{\sigma}^a \beta
\end{equation}
For standard model particles:\\
In the notation with left-handed and right-handed Weyl spinors one generation of SM fermion is
\begin{equation}
l\equiv\left(\begin{array}{c}\nu_L\\ e_L\end{array}\right)\quad e_R\quad \qquad\qquad q\equiv\left(\begin{array}{c}u_L\\ d_L\end{array}\right)\quad u_R\quad d_R
\end{equation}
In our notation, only left-handed spinors appear.
\begin{eqnarray}
&&\nu\equiv\nu_L\qquad e\equiv e_L\qquad \bar{e}\equiv -i\sigma^2 e_R^\ast\qquad \\
&&u\equiv u_L\qquad d\equiv d_L\qquad \bar{u}\equiv -i\sigma^2 u_R^\ast\qquad \bar{d}\equiv -i\sigma^2 d_R^\ast
\end{eqnarray}
And therefore for 1-generation of fermion
\begin{equation}
l\equiv\left(\begin{array}{c}\nu\\ e\end{array}\right)\quad \bar{e}\quad \qquad\qquad q\equiv\left(\begin{array}{c}u\\ d\end{array}\right)\quad \bar{u}\quad \bar{d}
\end{equation}
The notation of the SM gauge fields:\\
A frequently  used notation for gauge field
\begin{equation}
D_a=\partial_a + ig_s G_a^i T^i_s+ig W^i_a T^i +i\frac{g'}{2}B_aY \qquad\qquad tr(T_s^i T_s^j)=\frac{1}{2}\delta_{ij}\qquad\qquad tr(T^i T^j)=\frac{1}{2}\delta_{ij} \label{frequent}
\end{equation}
And the Lagrangian of the gauge field is
\begin{equation}
\mathcal{L}_{gauge}=-\frac{1}{4} G_{ab}^i G^{abi}-\frac{1}{4} W_{ab}^i W^{abi}-\frac{1}{4} B_{ab} B^{ab}
\end{equation}
Do the following switch to get our notation:
\begin{eqnarray}
&&\frac{g'}{2}B_a\to B_a \nn\\
&&gW^i_aT^i\to W_a \\
&&gG^i_aT_s^i\to G_a \nn
\end{eqnarray}
In our notation
\begin{eqnarray}
D_a&=&\partial_a +iG_a+iW_a+iB_a Y \nn\\
\mathcal{L}_{gauge}&=&-\frac{1}{2g_s^2}tr(G_{ab}G^{ab})-\frac{1}{2g^2}tr(W_{ab}W^{ab})-\frac{1}{g'^2}B_{ab}B^{ab} \nn\\
G_{ab}&=&\partial_a G_b-\partial_b G_a +i[G_a,G_b] \\
W_{ab}&=&\partial_a W_b-\partial_b W_a +i[W_a,W_b] \nn\\
B_{ab}&=&\partial_a B_b- \partial_b B_a  \nn
\end{eqnarray}
\qquad\qquad\qquad\qquad\qquad\qquad\qquad
\begin{tabular}{|c|c|}
\hline
 & $SU(3)_c\times SU(2)_L\times U(1)_Y$\\
 \hline
$q$&     $(3,2,\frac{1}{3})$      \\
$\bar{u}$& $(\bar{3},1,-\frac{4}{3})$\\
$\bar{d}$& $(\bar{3},1,\frac{2}{3})$      \\
\hline
$l$&    $(1,2,-1)$                              \\
$\bar{e}$&    $(1,1,2)$                          \\
$\bar{\nu}$&       $(1,1,0)$                    \\
\hline
$H$&       $(1,2,1)$                \\
$\tilde{H}$&        $ (1,2,-1)$                      \\
\hline
\end{tabular}\\
\subsection{The changes of anomaly term under fermion phase rotations, Fujikawa's result}\label{Fujikawa}
As we mainly deal with chiral gauge fields $W_a$ and $B_a$, Fujikawa's paper \cite{Fujikawa:1983bg} is the right reference. I just summarize some useful results in our notation.\\
For a left-handed Weyl spinor which couples to gauge field $W_a$ and $B_a$ like
\begin{equation}
q^\dagger i\bar{\sigma}^a(\partial_a+iW_a+iB_aY)q
\end{equation}
the phase rotation
\begin{equation}
q\to e^{i\alpha(x)} q
\end{equation}
will result in a change of anomaly term due to the change of path integral measure
\begin{equation}
\delta\mathcal{L}=\frac{\alpha(x)}{16\pi^2}[\,tr(W_{ab}\tilde{W}^{ab})+B_{ab}\tilde{B}^{ab}Y^2]\qquad\qquad \tilde{B}^{ab}\equiv\frac{1}{2}\epsilon^{abcd}B_{cd}
\end{equation}
In the SM Lagrangian  with $N_f$ families of fermions [Eqn.(\ref{5})],
when making the following local vector rotations:
\begin{eqnarray}
l&\to& e^{i\theta_1(x)}l \nn\\
\bar{e}&\to& e^{-i\theta_1(x)}\bar{e} \nn\\
q&\to& e^{i\theta_2(x)}q\\
\bar{u}&\to& e^{-i\theta_2(x)}\bar{u} \nn\\
\bar{d}&\to& e^{-i\theta_2(x)}\bar{d} \nn
\end{eqnarray}
(1)  $\mathcal{L}_{gauge}$, $\mathcal{L}_{Yukawa}$ and $\mathcal{L}_{Higgs}$ are  invariant.\\
(2) $\mathcal{L}_{M}$ has the following changes
\begin{equation}
\frac{g^2_\nu}{2M}[(\tilde{H}^\dagger l)(\tilde{H}^\dagger l)-(l^\dagger\tilde{H})(l^\dagger\tilde{H}) ]\to \frac{g^2_\nu}{2M}[e^{2i\theta_1(x)}(\tilde{H}^\dagger l)(\tilde{H}^\dagger l)-e^{-2i\theta_1(x)}(l^\dagger\tilde{H})(l^\dagger\tilde{H}) ]
\end{equation}
(3)  The change of $\mathcal{L}_{kinetic}$ is
\begin{equation}
\delta\mathcal{L}_{kinetic}=-(\partial_a \theta_1) l^\dagger \bar{\sigma}^a  l+(\partial_a \theta_1)\bar{e}^\dagger \bar{\sigma}^a  \bar{e}
                            -(\partial_a \theta_2) q^\dagger \bar{\sigma}^a  q+(\partial_a \theta_2) \bar{u}^\dagger \bar{\sigma}^a \bar{u}+(\partial_a \theta_2) \bar{d}^\dagger \bar{\sigma}^a  \bar{d}
\end{equation}
When $\partial_a \theta_1=(\delta_l,0,0,0)$ and $\partial_a \theta_2=(\delta_q,0,0,0)$ we get terms like
\begin{equation}
-\delta_l l^\dagger l\qquad\qquad -\delta_q q^\dagger q
\end{equation}
In Appendix \ref{8} we will explain that these terms cause energy shifts to fermions.\\
(4) Vector rotation does not give anomaly term to QCD gauge field $G_a$, but there will be changes of anomaly terms for chiral gauge fields $W_a$ and $B_a$ (due to path integral measure).
\begin{eqnarray}
\delta\mathcal{L}_{anomaly}&=& N_f(\theta_1+3\theta_2)\frac{1}{16\pi^2}  tr(W_{ab}\tilde{W}^{ab})\nonumber\\
 &&+ N_f\{\theta_1 [(-1)^2\times2 -2^2]+3\theta_2[(\frac{1}{3})^2\times2 -(-\frac{4}{3})^2-(\frac{2}{3})^2]\}\frac{1}{16\pi^2}  B_{ab}\tilde{B}^{ab}\nonumber\\
&=&N_f(\theta_1+3\theta_2)\frac{1}{16\pi^2}  [\,tr(W_{ab}\tilde{W}^{ab}) -2 B_{ab}\tilde{B}^{ab}]
\end{eqnarray}
Comparing to the frequently used notation discussed in Eq. (\ref{frequent})
\begin{equation}
\frac{1}{16\pi^2}  [\,tr(W_{ab}\tilde{W}^{ab}) -2 B_{ab}\tilde{B}^{ab}]\to \frac{1}{32\pi^2}[g^2W^i_{ab}\tilde{W}^{abi}-g'^2 B_{ab}\tilde{B}^{ab}]
\end{equation}
\subsection{Energy shifts and invariant quantities under basis changes}\label{8}
Here we provide some details about energy shifts in the context of massless left-handed Weyl fermion.
\subsubsection{Classical solutions for a free left-handed Weyl fermion without energy shift}
First consider the free theory
\begin{equation}
(A)\qquad\qquad\mathcal{L}=l^{\dagger}i\bar{\sigma}^a\partial_a l\qquad\qquad\label{10}
\end{equation}
Equation of motion
\begin{equation}
i\bar{\sigma}^a\partial_a l=0.
\end{equation}
Solutions with $p^a=(\vert\vec{p}\vert,\vec{p})$
\begin{equation}
l=u(p)e^{-ipx}\qquad\qquad l=v(p)e^{ipx}\qquad\qquad\textrm{with}\qquad\qquad \bar{\sigma}^a p_a u(p)=0\qquad\qquad \bar{\sigma}^a p_a v(p)=0
\end{equation}
For example with $E_0(\vec{p})=\vert\vec{p}\vert$, $u$ and $v$ are normalized such that
\begin{equation}
u(p)=v(p)=\sqrt{2E_0(\vec{p})}\left(\begin{array}{c}0\\1\end{array}\right)\qquad\qquad \textrm{when}\qquad\qquad p^a=(E_0(\vec{p}),0,0,E_0(\vec{p}))
\end{equation}
and
\begin{equation}
u(p)=v(p)=\sqrt{2E_0(\vec{p})}\left(\begin{array}{c}1\\0\end{array}\right)\qquad\qquad \textrm{when}\qquad\qquad p^a=(E_0(\vec{p}),0,0,-E_0(\vec{p}))
\end{equation}
The  general classical solution is
\begin{equation}
l=\int \frac{d^3p}{(2\pi)^3}\frac{1}{\sqrt{2E_0(\vec{p})}}(\alpha_{\vec{p}}u(p)e^{-ipx}+\beta^\ast_{\vec{p}}v(p)e^{ipx})
\end{equation}
\subsubsection{Basis changes make  $\delta_l l^\dagger l$ term appear and why it corresponds to an energy shift}
Do the change of basis $l\to e^{i\delta_lt}l$ on the free Lagrangian (\ref{10}), you will find
\begin{equation}
(B)\qquad\qquad\mathcal{L'}=l^{\dagger}i\bar{\sigma}^a\partial_a l-\delta_l l^\dagger l\qquad\qquad .
\end{equation}
The equation of motion
\begin{equation}
(i\bar{\sigma}^a\partial_a-\delta_l) l=0 .
\end{equation}
The classical solution of this equation can be obtained by the solution without energy shift times a factor $e^{-i\delta_l t}$, and it is consistent with the intuition that this Lagrangian is the free Lagrangian after basis change, and the solutions should be related by the similar transformation.
With $p^2=0$, the solutions are
\begin{equation}
l=u(p)e^{-ipx}e^{-i\delta_l t}\qquad\qquad l=v(p)e^{ipx}e^{-i\delta_l t}\qquad\textrm{with}\qquad \bar{\sigma}^a p_a u(p)=0\qquad\qquad \bar{\sigma}^a p_a v(p)=0.
\end{equation}
The  general classical solution is
\begin{equation}
l=\int \frac{d^3p}{(2\pi)^3}\frac{1}{\sqrt{2E_0(\vec{p})}}(\alpha_{\vec{p}}u(p)e^{-ipx}e^{-i\delta_l t}+\beta^\ast_{\vec{p}}v(p)e^{ipx}e^{-i\delta_l t})
\end{equation}
Therefore, with $E_0(\vec{p})=\vert\vec{p}\vert$
\begin{eqnarray}
E(\vec{p})&=&E_0(\vec{p})+\delta_l\qquad\qquad (\textrm{particles})\\
E(\vec{p})&=&E_0(\vec{p})-\delta_l\qquad\qquad (\textrm{antiparticles})
\end{eqnarray}
This is why we could interpret $-\delta_l l^\dagger l$ as an energy shift which shifts the energy of particles and antiparticles in opposite directions by the same amount.
\subsubsection{Quantization with creation and annihilation operators}\label{9}
For the theory without an energy shift
\begin{equation}
\mathcal{L}=l^{\dagger}i\bar{\sigma}^a\partial_a l\qquad\Longrightarrow\qquad \frac{\partial\mathcal{L}}{\partial\dot{l}}=il^\dagger
\end{equation}
Hamiltonian density
\begin{equation}
\mathcal{H}=l^\dagger i(\vec{\sigma}\cdot\vec{\nabla})l
\end{equation}
Quantize $l$ field by
\begin{equation}
l=\int \frac{d^3p}{(2\pi)^3}\frac{1}{\sqrt{2E_0(\vec{p})}}(a_{\vec{p}}u(p)e^{-ipx}+b^\dagger_{\vec{p}}v(p)e^{ipx})
\end{equation}
with the commutation relations
\begin{equation}
\{a_{\vec{p}},a^\dagger_{\vec{q}}\}=(2\pi)^3\delta^3(\vec{p}-\vec{q})\qquad\qquad \{b_{\vec{p}},b^\dagger_{\vec{q}}\}=(2\pi)^3\delta^3(\vec{p}-\vec{q})
\end{equation}
One can work out that the Hamiltonian (after dropping an infinite constant) is
\begin{equation}
H_0\equiv\int d^3x l^\dagger (i\vec{\sigma}\cdot\vec{\nabla})l=\int \frac{d^3p}{(2\pi)^3} E_0(\vec{p}) (a^\dagger_{\vec{p}}a_{\vec{p}}+b^\dagger_{\vec{p}}b_{\vec{p}})
\end{equation}
\subsubsection{Quantization of fermion with $\delta_l l^\dagger l$ term }
For theory with an energy shift
\begin{equation}
\mathcal{L'}=l^{\dagger}i\bar{\sigma}^a\partial_a l -\delta_l l^\dagger l\qquad\Longrightarrow\qquad \frac{\partial\mathcal{L'}}{\partial\dot{l}}=il^\dagger
\end{equation}
the Hamiltonian density is given by
\begin{equation}
\mathcal{H}=l^\dagger i(\vec{\sigma}\cdot\vec{\nabla})l+\delta_l l^\dagger l=\mathcal{H}_0+\mathcal{H}_\delta .
\end{equation}
The quantized $l$ field is given by
\begin{equation}
l=\int \frac{d^3p}{(2\pi)^3}\frac{1}{\sqrt{2E_0(\vec{p})}}(a_{\vec{p}}u(p)e^{-ipx}e^{-i\delta_l t}+b^\dagger_{\vec{p}}v(p)e^{ipx}e^{-i\delta_l t})
\end{equation}
where the operators $a$, $a^\dagger$, $b$, $b^\dagger$, $u$ and $v$ have the same property as described in \ref{9}.
One can work out the Hamiltonian (after dropping an infinite constant) and we find
\begin{eqnarray}
H_0&\equiv&\int d^3x l^\dagger (i\vec{\sigma}\cdot\vec{\nabla})l=\int \frac{d^3p}{(2\pi)^3} E_0(\vec{p}) (a^\dagger_{\vec{p}}a_{\vec{p}}+b^\dagger_{\vec{p}}b_{\vec{p}})\\
H_\delta&\equiv&\int d^3x\,\, \delta_l l^\dagger l=\int \frac{d^3p}{(2\pi)^3} \delta_l (a^\dagger_{\vec{p}}a_{\vec{p}}-b^\dagger_{\vec{p}}b_{\vec{p}}) .
\end{eqnarray}
Again, we recover the energy shift explanation in the context of the quantized theory.
\subsubsection{Discrete symmetries}
We remark that the operator $\delta_l l^\dagger l$ is even under a $T$ (time reversal) transformation, and odd under $CP$:
\begin{eqnarray}
(T)\qquad\qquad \delta_l l^\dagger l&\to& +\delta_l l^\dagger l \nn\\
(CP)\qquad\qquad \delta_l l^\dagger l&\to& -\delta_l l^\dagger l\\
(CPT)\qquad\qquad \delta_l l^\dagger l&\to& -\delta_l l^\dagger l \nn
\end{eqnarray}
At the operator level, it may be seen by looking at the transformation on operators
\begin{eqnarray}
(T)\qquad\qquad a_{\vec{p}}\to a_{-\vec{p}}\qquad\qquad b_{\vec{p}}\to b_{-\vec{p}}   \\
(CP)\qquad\qquad a_{\vec{p}}\to b_{-\vec{p}}\qquad\qquad b_{\vec{p}}\to a_{-\vec{p}}  \nn
\end{eqnarray}

\subsection{Energy-momentum tensor in a time dependent background from Noether's theorem}\label{Noether}
\subsubsection{Noether's theorem and energy-momentum tensor in a background}
Consider a general Lagrangian, $\mathcal{L}(\phi,\partial_a\phi,x^a)$, and allow it to depend on $x^a$ explicitly, so that it may apply to theories with time dependent background fields. We will take the partial derivative of the Lagrangian with respect to $x^a$, and we use the following two quantities for different meanings
\begin{equation}
\frac{\partial\mathcal{L}}{\partial x^a}\ne \partial_a\mathcal{L}
\end{equation}
The one on the LHS is the partial derivative which keeps $\phi$ and $\partial_{a}\phi$ fixed, while the one on the rhs is
\begin{equation}
\partial_a\mathcal{L}\equiv \frac{\partial\mathcal{L}}{\partial x^a}+\frac{\partial\mathcal{L}}{\partial \phi}\partial_a\phi+\frac{\partial\mathcal{L}}{\partial(\partial_b \phi)}\partial_a(\partial_b\phi) . \label{define}
\end{equation}
Given the action
\begin{equation}
S=\int d^4x \mathcal{L}(\phi,\partial_a\phi,x^a),
\end{equation}
take infinitesimal variation $\delta\phi(x)$ which vanishes at the boundary, and then integrate by parts
\begin{eqnarray}
\delta S&=&\int d^4x [\frac{\partial\mathcal{L}}{\partial \phi}\delta\phi+\frac{\partial\mathcal{L}}{ \partial(\partial_a\phi)}\delta(\partial_a\phi)]\nonumber\\
&=&\int d^4x \,\,\delta\phi[\frac{\partial\mathcal{L}}{\partial \phi}-\partial_a\big(\frac{\partial\mathcal{L}}{ \partial(\partial_a\phi)}\big)].
\end{eqnarray}
Therefore, in the case the Lagrangian depends on $x^a$ explicitly, we are still be able to derive the Euler-Lagrange equation
\begin{equation}
\frac{\partial\mathcal{L}}{\partial \phi}-\partial_a\big(\frac{\partial\mathcal{L}}{ \partial(\partial_a\phi)}\big)=0 .
\end{equation}
Consider a constant infinitesimal space-time translation $\epsilon^b$. Using Eq. (\ref{define}) we find
\begin{equation}
\epsilon^b\frac{\partial\mathcal{L}}{\partial x^b}+\frac{\partial\mathcal{L}}{\partial \phi}(\epsilon^b\partial_b\phi)+\frac{\partial\mathcal{L}}{\partial(\partial_a \phi)}\epsilon^b\partial_b(\partial_a\phi)=\epsilon^b\partial_b\mathcal{L}.
\end{equation}
Then, use the Euler-Lagrangian equations to derive
\begin{equation}
\epsilon^b\partial_a[\big(\frac{\partial{\mathcal{L}}}{\partial(\partial_a\phi)}\big)\partial_b\phi-\delta^a_{\,\,\,b}\mathcal{L}]=-\epsilon^b\frac{\partial\mathcal{L}}{\partial x^b}.
\end{equation}
It is valid for any $\epsilon^b$, and let us define the energy-momentum tensor to be
\begin{equation}
T^{a}_{\,\,\,b}\equiv\big(\frac{\partial{\mathcal{L}}}{\partial(\partial_a\phi)}\big)\partial_b\phi-\delta^a_{\,\,\,b}\mathcal{L}\qquad\Longrightarrow\qquad \partial_a T^a_{\,\,\,b}=-\frac{\partial\mathcal{L}}{\partial x^b}. \label{conservation}
\end{equation}
Therefore, if the Lagrangian does not explicitly depend on $x^a$, we will find $\partial_a T^a_{\,\,\,b}=0$ and the energy-momentum tensor is conserved. On the other hand, if the Lagrangian explicitly depends on $x^a$, the energy-momentum tensor is not conserved.
\subsubsection{The energy nonconservation in basis (A) of the toy model}
Recall that the Lagrangian for the toy model in basis (A) is (with $\partial_a\theta=(\delta,0,0,0)$)
\begin{equation}
(A)\qquad\mathcal{L}=l^\dagger i\bar{\sigma}^a D_a l+ \sum_{i=1}^3 q_i^\dagger i\bar{\sigma}^aD_a q_i-\frac{1}{2g^2}tr(W_{ab}W^{ab})-\frac{\theta(x)}{16\pi^2}tr(W_{ab}\tilde{W}^{ab})
\end{equation}
Notice that the Lagrangian depends on $x^a$ explicitly only through the background $\theta(x)$. Thus using Eq. (\ref{conservation}) we find\footnote{In our case $\delta=\textrm{const}$, it is possible to define the energy-momentum tensor another way and make it a conserved tensor. It is because when $\delta=\textrm{const}$ the $\theta(x)W\tilde{W}$  is equivalent to a time independent term up to a total derivative.
For a general time dependent background it is not possible to define a conserved energy-momentum tensor.}
\begin{equation}
\partial_a T^a_{\,\,\,b}=\delta^0_b \frac{\delta}{16\pi^2}tr(W_{ab}\tilde{W}^{ab})
\end{equation}
$\partial_a T^a_{\,\,\,i}=0$ with $i=1,2,3$ tells you 3-momentum is conserved, and $\partial_a T^a_{\,\,\,0}\ne0$ tells you the energy is not conserved. [If $\theta(x)$ depends on the 3-dimensional space, we expect the 3-momentum not to be conserved.] The energy of the system is
\begin{equation}
E(t)=\int d^3x T^0_{\,\,\,0}(\vec{x},t)
\end{equation}
The amount of energy nonconservation is
\begin{equation}
E(t_2)-E(t_1)=\int _{t_1}^{t_2}dt\int d^3x \,\,\partial_a T^a_{\,\,\,0}=\int _{t_1}^{t_2}dt\int d^3x \,\,\frac{\delta}{16\pi^2}tr(W_{ab}\tilde{W}^{ab})
\end{equation}
The instanton number
\begin{equation}
\nu\equiv\int d^4x\,\, \frac{1}{16\pi^2}tr(W_{ab}\tilde{W}^{ab})
\end{equation}
Therefore, the change of energy is $+\nu\delta$ for an instanton process. This confirms our result in Section \ref{Boltzmann}. In other words, in the anomalous interaction
\begin{equation}
l+q_1\rightleftharpoons \bar{q}_2+\bar{q}_3
\end{equation}
the energy is not conserved from the viewpoint of basis (A) and
\begin{equation}
E_{\bar{q}_2}+E_{\bar{q}_3}=E_l+E_{q_1}+\delta
\end{equation}

\bibliographystyle{utphys}
\bibliography{bibfile}

\end{document}